\documentclass[aps,prl,reprint,groupedaddress]{revtex4-1}
\usepackage{graphicx}
\usepackage{dcolumn}
\usepackage{bm}

\begin{document}

\preprint{AIP/123-QED}

\title[]{Anisotropic magneto-Coulomb effect versus spin accumulation in a ferromagnetic single-electron device}
\author{A. Bernand-Mantel}
 \altaffiliation[Present address ]{Institut N\'eel, CNRS et Universit\'e Joseph Fourier, BP 166, F-38042 Grenoble Cedex 9, France
}
  \email{anne.bernand-mantel@grenoble.cnrs.fr}
\author{P. Seneor}%
\author{K. Bouzehouane}%
\author{S. Fusil}%
\author{C. Deranlot}%
\author{F. Petroff}%
\author{A. Fert}%
\affiliation{ 
Unit\'e Mixte de Physique CNRS/Thales and Universit\'e Paris Sud 11, 91767 Palaiseau Cedex, France
}%

\date{\today}

\begin{abstract}
We investigate the magneto-transport characteristics of nanospintronics single-electron devices. The devices consist of single non-magnetic nano-objects (nanometer size nanoparticles of Al or Cu) connected to Co ferromagnetic leads. The comparison with simulations allows us attribute the observed magnetoresistance to either spin accumulation or anisotropic magneto-Coulomb effect (AMC), two effects with very different origins. The fact that the two effects are observed in similar samples demonstrates that a careful analysis of Coulomb blockade and magnetoresistance behaviors is necessary in order to discriminate them in magnetic single-electron devices. As a tool for further studies, we propose a simple way to determine if spin transport or AMC effect dominates from the Coulomb blockade I-V curves of the spintronics device. 
\end{abstract}

\pacs{Valid PACS appear here}
\maketitle
\begin{figure}
\includegraphics{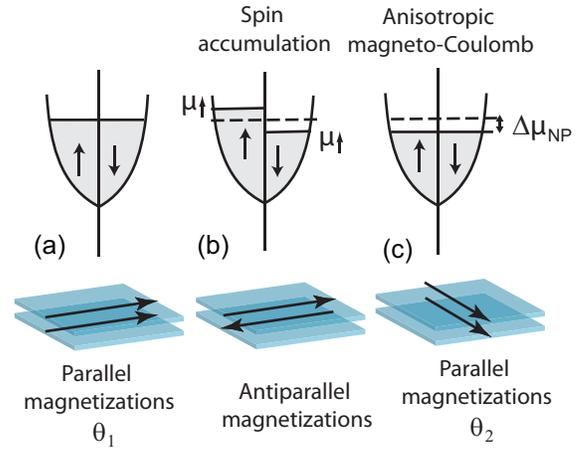}
\caption{\label{fig:dos} Schematic representation of the two Co electrodes magnetization (bottom)  and corresponding nanoparticle density of states (top) in the case  of (a) parallel magnetizations (b) Antiparallel magnetizations inducing a spin accumulation on the nanoparticle (c) AMC effect induced by a rotation of the magnetizations}
\end{figure}

One current direction of spintronics consists in studying spin dependent transport in a single nano-object connected to ferromagnetic leads, i.e. nanospintronics \cite{Seneor07}. In this type of system the ultra small capacitance between the electrodes and the central island gives rise to the Coulomb blockade effect, which allows a precise control of the electron transport through the nano-object. The interplay between this single electron transport and spin dependent transport has been the subject of many theoretical studies in the past ten years \cite{Barnas08}. On the experimental side the number of studies has been strongly limited by the numerous difficulties encountered when trying to connect a single nano-object to ferromagnetic leads \cite{Seneor07}. In line with the theoretical expectations, in the few successful works, the observation of Coulomb blockade together with a magnetoresistance  signal has been systematically attributed to spin dependent transport leaving aside the effect of the electrodes \cite{Sahoo05,Hamaya07,Yakushiji05,Wei07,ABM06,Hamaya08,Merchant08,Davidovic09}. However, some recent experiments have pointed out that in Coulomb blockade systems a very similar MR signal could be induced by an anisotropic magneto-Coulomb (AMC) effect \cite{Wunderlich06,Liu08,ABM09,Zwanenburg9,cleuziou11}. It was demonstrated that the magnetic anisotropy of the electrode can induce a variation of its electrochemical potential during magnetization rotation \cite{ABM09}. Importantly, one can see that almost similar MR curves can lead to very different analysis depending on the chosen interpretation : spin accumulation or AMC. Therefore, while easily confused, both effects should not be mistaken.\\
The aim of this letter is to shed some light on this issue by reporting and comparing both effects obtained in similar samples. 
In order to do so,  devices containing single non-magnetic nanoparticles connected to ferromagnetic electrodes have been measured. 
In previous works, we reported separately on the observation of  spin accumulation \cite{ ABM06, Seneor07}  and  AMC effect  \cite{ ABM09} in this type of devices. 
In this paper, two samples showing representative features of each effects are presented and compared.
The principal differences between AMC and spin accumulation effects are highlighted  and  their  occurrence  in the devices is discussed. \\
\begin{figure}
\includegraphics{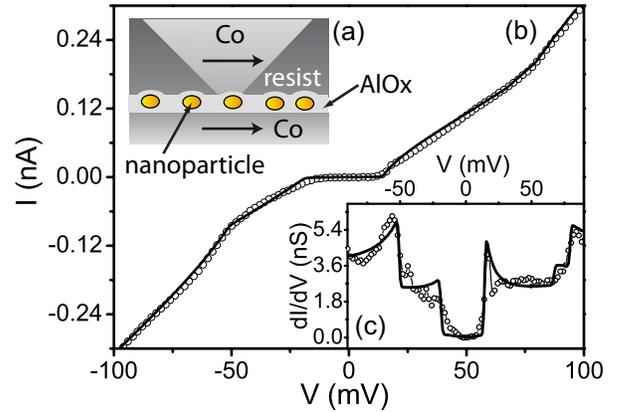}
\caption{\label{fig:CB} (a) Schematic representation of the ferromagnetic single electron device (b) I-V measured at 1.5 K on sample S1 (circles) containing Cu nanoparticles and single nanoparticle Coulomb blockade simulation (line) with C$_{1}$=1.86 aF, C$_{2}$=2.46 aF, R$_{1}$=100 M$\Omega$, R$_{2}$=150 M$\Omega$ and Q$_{0}$=0.28 e. (c) dI/dV of the curves in (b)}
\end{figure}
These samples consist respectively of a single Cu (sample S1) or Al (sample S2) nanoparticle. 
The elaboration process is based on a nanoindentation technique which is used to fabricate a sub-10nm nanocontact on top of a nanoparticle assembly (see Fig.~\ref{fig:CB}(a)). This process allows the connection of a single nanoparticle of few nanometers in diameter \cite{ABM10}. The structure, grown by sputtering, starts with 15 nm of Co covered with 0.6 nm of Al further oxidized in 50 Torr of O$_{2}$ to form a tunnel barrier. The nanoparticle assemblies are then obtained by the deposition of 1.5 nm of Cu (S1) or 2 nm of Al (S2). The second alumina barrier is made by depositing 0.6 nm of Al further oxidized in 50 Torr of O$_{2}$ for S1 and oxidation of the outer shell of the Al nanoparticles in 1.10$^{-2}$ Torr of O$_{2}$ for S2. After the nanoindentation process, the nano-hole is filled with 10 nm of Co and 190 nm of Au to form the top electrode and finalize the nanocontact. A schematic representation of the device is shown in Fig.~\ref{fig:CB}(a).\\
\begin{figure}
\includegraphics{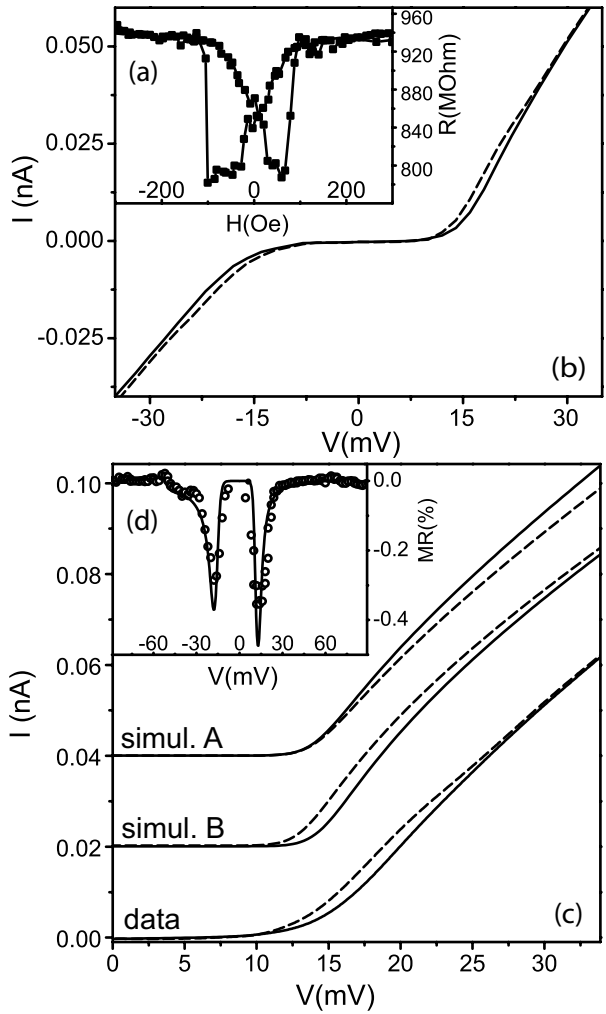}
\caption{\label{fig:S1} (a) Resistance versus magnetic field of sample S1 measured at V=20 mV  (b) I-Vs measured at 1.5 K on sample S1 and recorded at magnetic fields corresponding to high resistance I$_{H}$ (line, 380 Oe) and low resistance I$_{L}$ (dash, -60 Oe) states. (c) Comparison of the experimental curves indicated as "data" and simulations. For the simulations the resistance and capacitances parameters are those given in Fig.~\ref{fig:CB}. Simul. A : spin accumulation with $\delta$=1me V, $P$=0.35, $\tau_{sf}$=20ns. Simul. B : AMC with Q$_{0}$=0.28 e, Q$_{0}$'=0.298 e. (d) Magnetoresistance MR=(I$_{H}$-I$_{L}$)/I$_{L}$ versus voltage calculated from experimental (circles) and simulated (line) I-Vs}
\end{figure}
Transport measurements were carried out in a cryostat at 1.5 K. A first current-voltage I-V characteristic measured on sample S1 is shown in Fig.~\ref{fig:CB}(b). This curve presents typical features of Coulomb blockade : suppression of the current at low voltage and steps in the current at higher voltage (visible as peaks in the dI/dV in Fig.~\ref{fig:CB}(c)). We also performed simulations (see Fig.~\ref{fig:CB} (b,c)) to extract the system parameters following the method described in ref \cite{ABM10}. The good agreement between simulations (black line) and experimental data (circles) shows that  transport occurs through a single isolated nanoparticle. We now focus on the magnetic properties of the system. In Fig.~\ref{fig:S1}(a) we present resistance as a function of magnetic field measured at a voltage of 20 mV. At low field, an hysteretic magnetoresistance (MR) signal with a maximum amplitude of 20\% is observed. In Fig.~\ref{fig:S1}(b) we show two I-V curves recorded at two different magnetic fields corresponding to the high resistance (line, $H_1=380\ Oe$) and low resistance (dash, $H_2=-60\ Oe$) states of the MR curve.\\
 We first discuss the origin of this observed MR based on a spin dependent transport model. Our system is equivalent to a double tunnel junction Co/Al$_{2}$O$_{3}$/Cu/Al$_{2}$O$_{3}$/Co in which each junction possesses one ferromagnetic (Co) and one nonmagnetic (Cu) electrode. If we consider the classical Julliere's model \cite{Juliere75} where the TMR=2P$_{1}$P$_{2}$/(1-P$_{1}$P$_{2}$) is proportional to the electrodes spin polarizations, we do not expect any tunneling magnetoresistance (TMR) in our system as P$_{Cu}$=0. However the spin polarized current which is injected from the first electrode can conserve its spin polarization while passing through the nanoparticle if the spin flip rate on the nanoparticle is smaller than the electron injection rate $\Gamma_{sf}=1/\tau_{sf}<\Gamma_{inj}=I/e$. In that case an accumulation of one spin direction and a depletion of the other will build up for antiparallel magnetizations, giving rise to a splitting $\Delta\mu$=$\mu_{\uparrow}-\mu_{\downarrow}$ between spin up and spin down electrochemical potentials  on the nanoparticle (see Fig.~\ref{fig:dos}(b)). On the contrary no spin accumulation is expected for parallel magnetizations and the nanoparticle electrochemicals potentials remain unchanged (Fig.~\ref{fig:dos}(a)). 
 In the antiparallel case the non equilibrium spin accumulation is proportional to the spin lifetime on the nanoparticle $\tau_{sf}$, the electrodes spin polarization $P$ , the current $I$ and the nanoparticle energy level spacing $\delta$ : $\Delta\mu$=$\tau_{sf}$PI$\delta/e$. 
This spin accumulation induces  a finite TMR effect which culminates to P$^{2}$/(1-P$^{2}$) if we consider a symmetric junction and a infinite spin lifetime on the nanoparticle  \cite{Braatas99}.
 To check whether our results could be interpreted by this effect we have included a spin accumulation term $\Delta\mu$ in our Coulomb blockade simulations following the work from Brataas et al.  \cite{Braatas99}. The result is indicated as Simul.A in Fig.~\ref{fig:S1}(c) where we have represented the I-V curves for parallel (continuous line) and antiparallel (dashed line) magnetizations. 
  We emphasize that the difference between the two simulated I-V curves is minimum at low current and maximum a high current. This behavior is expected from the spin accumulation model as $\Delta\mu$ is proportional to the current. In addition the sign of the MR is positive. Those two characteristics are in opposition to what is observed experimentally. This can be seen by comparing the simulated curves with the experimental ones indicated as data in Fig.~\ref{fig:S1}(c). On the experimental data the MR is negative and the difference between  the two I-V curves is maximum at low current and decreases with the current. Those differences suggest that the MR might have another origin than spin accumulation in this sample. \\
We now discuss an explanation of the MR in terms of AMC effect. On the experimental I-V curves in Fig.~\ref{fig:S1}(c) one can see that the low resistance state curve presents a reduced Coulomb gap as compared to the high resistance state. Such shift of the threshold voltage is characteristic of the AMC effect induced by the anisotropy of the Co electrode, as demonstrated in a previous work \cite{ABM09}. In the Co electrode the spin-orbit coupling links the electronic structure and the magnetization. Due to this coupling the magnetic anisotropy gives rise to an anisotropy of the electrochemical potential of the magnetic electrode $\mu_{F}$. As a consequence $\mu_{F}$ depends on the magnetization direction $(\theta)$. The two I-Vs of  Fig.~\ref{fig:CB}(b) measured at $H_1$ and $H_2$  correspond to two different magnetization angles $\theta_{1}$ and $\theta_{2}$. This is due to the rotation of the magnetization during its reversal. The electrochemical potential variation  $\Delta \mu_{F}$= $\mu_{F}(\theta_{1})$-$\mu_{F}(\theta_{2})$ of the electrode induces a variation of the nanoparticle's electrochemical potential $\Delta \mu_{NP}$ \cite{ABM09} as represented in Fig.~\ref{fig:dos}(c). This AMC effect can be simulated by introducing a variation $\Delta \mu_{NP}$ in our Coulomb blockade model. The result of this simulation is indicated as Simul.B in Fig.~\ref{fig:S1}(c). The experimental and simulated MR versus V are also compared in Fig.~\ref{fig:S1}(d). The best agreement between simulated and experimental I-V curves is obtained for $\Delta \mu_{NP}$=$e\Delta Q/C_{tot}$=0.7 meV.  The AMC simulation reproduces very well the observed effect in contrast to the spin accumulation model. This indicates that the spin accumulation effect is negligible in this sample and that the MR is dominated by the AMC effect. As discussed in our previous work, \cite{ABM09} the observed energy variation in the meV range is large compared to what is expected for bulk Co. This is due to the very local character of our measurement which probes the anisotropy at a single grain scale (few nm) where the anisotropy is enhanced by the reduced local symmetries at the grain surface. \\
\begin{figure}
\includegraphics{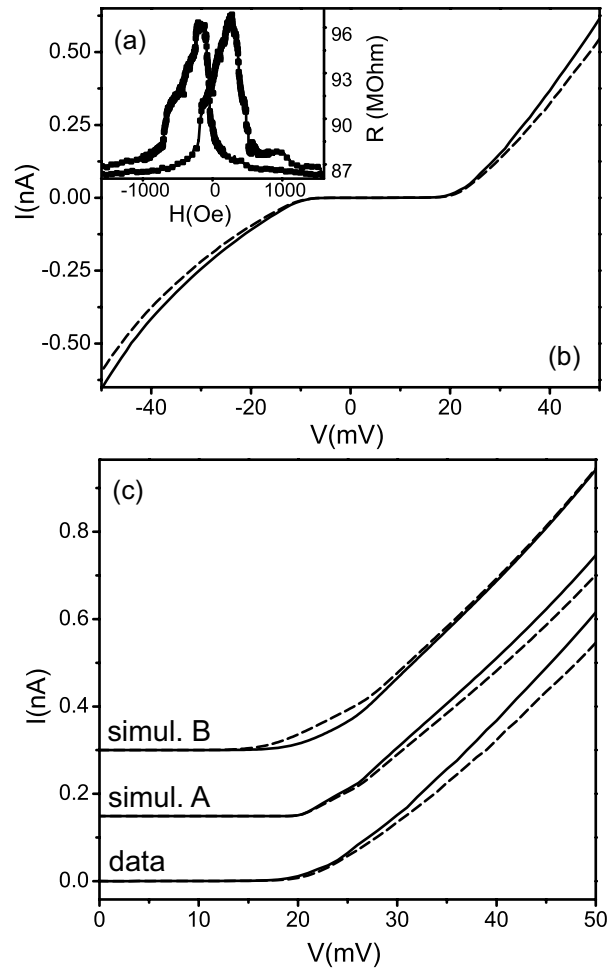}
\caption{\label{fig:S2} (a) Resistance versus magnetic field of sample S2 containing a Al nanoparicle, taken at V=50 mV  (b) I-Vs measured at 1.5 K on sample S2 and taken at magnetic fields corresponding to low resistance (parallel) (line, 1000 Oe) and high resistance (anti-parallel)  (dash, -150 Oe) states.(c) Comparison of the experimental curves indicated as "data" and simulations. Simulation parameters : C$_{1}$=4.7 aF, C$_{2}$=1.8 aF, R$_{1}$=20 M$\Omega$, R$_{2}$=60 M$\Omega$,  Q$_{0}$=0.27 e, Simul. A : spin accumulation with $\delta$=1me V, $P$=0.35 and $\tau_{sf}$=5 ns, Simul. B : AMC with Q$_{0}$=0.27 e, Q$_{0}$'=0.3 e}
\end{figure}
We now focus on a second sample S2 where the opposite behavior has been observed.  It contains an aluminum nanoparticle connected to Co leads. A resistance versus magnetic field measurement taken at a voltage of 50 mV and two I-V curves corresponding to the high and low resistance states of the MR curve are showed in Fig.~\ref{fig:S2}(a) and (b). Those results present similarities to what is expected in the case of spin accumulation effect : the TMR is positive and the effect is increasing with current. To confirm this we compare on Fig.~\ref{fig:S2}(c) the experimental I-Vs with Coulomb blockade simulations including spin accumulation. For this sample, the observed effect is well reproduced by taking into account only the spin accumulation effect (Simul.A on Fig.~\ref{fig:S2}(c)). In contrast, no threshold voltage shift is observed indicating that the AMC effect is negligible in this particular sample.  A simulation corresponding to the AMC effect, showing the expected threshold voltage shift, is presented in Fig.~\ref{fig:S2}(c) (Simul. B) for comparison.
The amplitude of the observed spin lifetime $\tau_{sf}$ can be estimated from a comparison between the experiment and the spin accumulation simulation to be  5 ns which is 50 times larger than the $\tau_{sf}$ measured in polycrystalline thin Al films \cite{Jedema02}. Other experiments have reported such a spin lifetime enhancement in metallic nanoparticles \cite{Yakushiji05,Wei07,Seneor07}.  \\
 In the following, we discuss on the occurrence of the AMC and spin accumulation effects in our devices. The AMC effect manifest itself as an electrochemical potential shift $\Delta \mu_{F}$ on the electrodes. The impact of this shift ($\Delta \mu_{F}$) on the Coulomb blockade I-V characteristic depends on its ratio to the charging energy of the nano-object ($ E{c}=e^2/2C_{tot}$). This means that considering a ferromagnetic electrode such as Co, the AMC effect could be observed in nano-objects with a charging energy in the tens of meV range and below (such as Carbon nanotubes for example). Then, the difference between sample S1 and S2, for which the electrodes are identical, can be understood by the very local character of the measurement. The magnetoresistance is probed via the few nanometers nanoparticle, at a single grain scale on the polycrystalline Co electrode. If the local grain possesses an easy axis different from the applied magnetic field direction, the coherent rotation of the magnetization will give rise to AMC magnetoresistance. If on the contrary, the particular local grain possesses an easy axis aligned with the magnetic field, or a weaker anisotropy,  no AMC magnetoresistance will be observed. Concerning spin accumulation, its amplitude is mainly governed by two parameters : the current intensity (linked to the device resistances) and the spin lifetime  on the nanoparticle. This can be roughly summarized by saying that for a spin accumulation to build up on the nano-object, the spin injection/extraction rate $\Gamma_{inj}=I/e$ must be higher than the spin flip rate $\Gamma_{sf}=1/\tau_{sf}$ as discussed earlier.   For the samples presented in this letter, we see that this criteria  is more likely to be satisfied for sample  S2 than for sample S1. Indeed, the current of sample S2 is 1 order of magnitude larger than that of S1 and the spin lifetime  is expected to be larger in S2 due to the weaker spin-orbit coupling in Al compared to Cu. While in this particular case spin accumulation is observed for Al and not Cu nanoparticles,  it should also in principle be observable for Cu if the injected current is sufficiently strong.\\
In summary, we investigated magneto-transport in ferromagnetic single-electron devices. 
The comparison of the observed behaviors with simulations allowed us to demonstrate that either spin accumulation or AMC effects can be observed in similar samples.  The signature of the spin accumulation $\Delta \mu$ at low voltage is its linear increase with the current while the AMC electrochemical potential shift $\Delta \mu$ is constant with increasing current, leading to a fast decrease of the MR with voltage and a characteristic threshold voltage shift. While it is difficult to predict with certitude if AMC or spin accumulation will be dominating in a Coulomb blockade spintronics device, we showed that a careful analysis of Coulomb blockade and magnetoresistance behaviors allows to discriminate between the two effects. This should help the understanding and development of future nanospintronics devices.

 \begin{acknowledgments}
We acknowledge support from the French ANR under grant Chemispin and RTRA triangle de la physique and CNano Ile de France.
\end{acknowledgments}

%
\end{document}